\documentclass[a4paper,11pt]{article}
\usepackage{jinstpub} % for details on the use of the package, please see the JINST-author-manual
\usepackage{subcaption}
\usepackage{float}
\usepackage{lineno}
\usepackage{orcidlink}
\usepackage[compatibility=false]{caption}
\usepackage{ulem}
\title{\boldmath Experimental Characterization of Bulk Micromegas for Development of Active Target
Time Projection Chamber for Nuclear Astrophysics Studies}

\author[a,1]{Pralay Kumar Das\,\orcidlink{0000-0002-0810-3528}\note{Corresponding author},}
\author[b]{Supratik Mukhopadhyay\,\orcidlink{0000-0002-8040-2338},}
\author[a]{Nayana Majumdar\,\orcidlink{0000-0002-4178-3000}}
\affiliation[a]{Nuclear, Atomic \& Particle Physics Division, Saha Institute of Nuclear Physics, A CI of Homi Bhabha National Institute, 1/AF Block, Bidhannagar, Kolkata, West Bengal, India}
\affiliation[b]{Research Wing, Naihati Prolife, Naihati, West Bengal, India}
\emailAdd{pralay.das@saha.ac.in}

\abstract{A Micromegas-based active target Time Projection Chamber, namely Saha Active Target TPC (SAT-TPC), is under development at Saha Institute of Nuclear Physics for its application in a nuclear astrophysics experiment to measure the branching ratio of the direct and sequential decay of the Hoyle state of $^{12}$C. A bulk Micromegas was  characterized to optimize its operating parameters and its performance for $\alpha$-particle tracking was investigated using gas mixtures Ar + CO$_{2}$ (90:10) and Ar + i-C$_4$H$_{10}$ (95:5) at atmospheric pressure using 5.9~keV X-rays and 5.48~MeV $\alpha$-particles from radioactive sources.  A hydrodynamic transport model was used to simulate the charge-deposit profile of the $\alpha$-particle track on the Micromegas readout and compared to the experimental data to complement the experimental measurement as well as validate the numerical model. The simulated charge-deposit profile showed a qualitative consistency
with the experimental measurement, supporting the suitability of the
bulk Micromegas for the SAT-TPC prototype and the applicability of the
hydrodynamic model for describing the detector response.}

\keywords{Micromegas, Active-target TPC, Energy resolution, $\alpha$-tracking, Gas gain}

\begin{document}
\phantomsection
\maketitle
\flushbottom

\section{Introduction}
\label{sec:intro}

%The decay process of the Hoyle state of $^{12}C$ nucleus holds an important clue to understand the creation of carbon and heavier elements in helium burning and evolution of the stars. The cross-section of the sequential and direct decay branches has been under detailed investigation for last few decades and quite a large number of experiments have been accomplished to determine the branching ratio of the Hoyle state decay modes~\cite{Freer94, Manfredi12, Kirsebom12, Itoh14, Morelli16, Aquila17, Rana19, Bishop20, Smith20}. 
The active target Time Projection Chamber is an advanced variation of the standard Time Projection Chamber (TPC) technology~\cite{Nyg78}, distinguished by the unique dual functionality of its gas volume, which acts simultaneously as both the reaction target and the active detection medium. This defining feature offers a next-generation, highly sophisticated detection methodology, tailored for modern nuclear physics experiments, especially the nuclear reactions in inverse kinematics using radioactive ion beams. This device provides near-$4\pi$ coverage and enables efficient three-dimensional tracking of charged nuclei which lead to precise measurement of nuclear reaction or decay cross-sections through kinematic reconstruction of the event and identification of nuclear products utilizing their energy loss in the large gas volume. This methodology effectively mitigates several issues of conventional techniques that employ solid targets and solid-state detector arrays, most notably the severe energy straggling of low-energy reaction products within the target~\cite{Baz20},  while offering the convenience of setting up and operating a stand-alone compact device.

It is worth mentioning here that there are drawbacks as well in the implementation of the active target TPC in the nuclear physics experiments aimed at measuring low cross-sections. The energy straggling in the gaseous targets may be limited by using noble gas only, however, the lack of a quencher gas prevents the stopping of electron avalanches at the amplification stage, often leading to uncontrollable multiplication and sparking. So, the choice of active gas is critical for both satisfying the experimental objectives and achieving optimal TPC performance. 

Nevertheless, the advantages of the device have caused a paradigm shift in the experimental technique to explore wide range of research problems in the nuclear and nuclear astrophysics fields, spanning over shell evolution, cluster structure, exotic decays, giant resonances, fusion-fission, etc. Several examples of such applications of this novel device include MAYA~\cite{Demon07}, MUSIC~\cite{Avila17}, MAIKo~\cite{Furuno18}, ACTAR~\cite{Mauss19}, AT-TPC~\cite{Ayyad20}, ELITPC~\cite{Gai20}, TexAT~\cite{Koshchiy20}, MATE~\cite{Zhang21}, GADGET~\cite{Maha24}, etc. 
Such is the case of studying the extremely rare direct decay channel of the Hoyle state of $^{12}$C, that offers critical insight into the role of carbon in stellar nucleosynthesis. The use of active target TPC enabled high-precision determination of the direct decay branching ratio through direct (0.043\%)~\cite{Bishop20} and indirect (0.00057\%)~\cite{Smith20} approaches, employing Micromegas-based active target TPC and optical active target TPC, respectively, while the best limit with the silicon detectors, achieved so far, is 0.019\%~\cite{Rana19}.

%Micromegas (MICRO-MEsh GAseous Structure) detector, known for its excellent spatial resolution, high rate capability, and robustness, is well suited as position readout component in the AT-TPC. Their fine granularity and fast response make them ideal for precise tracking and energy-loss measurements in AT-TPC applications. A Micromegas manufactured using the bulk technique at CERN \cite{IG06} was used. The anode readout plane has a one-dimensional segmentation of 10 strips over an area of $10 \times 10~\mathrm{cm^2}$ and a $128~\mu\mathrm{m}$ amplification gap.

An R\&D work is in progress to build an active target TPC at Saha Institute of Nuclear Physics, India, namely, the Saha AT-TPC (SAT-TPC). The objective of the developmental work is to implement the device in measuring low cross-sections through precise reconstruction of the event kinematics, aided by the particle tracking in an almost background-free scenario. A comprehensive investigation into the decay mechanisms of the Hoyle state has been selected as the primary objective of the utilization of the SAT-TPC. This focus is driven by the fact that accurate determination of the direct channel branching ratio remains inconclusive, as indicated by the existing experimental results~\cite{Bishop20,Smith20,Rana19}. To initiate the R\&D work in a systematic way, we studied the design and operational guidelines of the SAT-TPC using numerical simulation, based on a hydrodynamic model~\cite{DasModel2025}. The performance of the device with the selected design parameters and operating conditions was investigated for optimization purpose and dealing with the beam-induced space charge by tuning the readout granularity and beam intensity. In addition, we developed a data analysis framework for classification of the direct and the sequential decay events, employing Convolutional Neural Network (CNN)~\cite{DasCNN2025} which would be useful in identifying the rare direct $3\alpha$-decay in presence of the dominant sequential decay and other interactions taking place due to use of mixed active target. The final SAT-TPC will require highly precise three-dimensional kinematic reconstruction to resolve subtle differences in the decay topologies. 

We have initiated the present work as a foundation hardware activity towards the fabrication of a SAT-TPC prototype. The primary objective of this work is the experimental benchmarking of the Micromegas prototype for its implementation in the SAT-TPC prototype as the end-plate readout along with the validation of the associated numerical simulation framework used for predicting the device performance and analyzing the experimental observations. A bulk-Micromegas~\cite{Gio06} prototype was  characterized with two active gas mixtures, Ar + CO$_2$ (90:10) and Ar + i-C$_4$H$_{10}$ (95:5), at atmospheric pressure, in a setup with a 4~cm long drift region, equipped with a field cage. The performance of the detector in tracking $\alpha$-particle of nearly similar energy of that produced in Hoyle state decay was studied using a very basic readout scheme. The results provide the operating field ratios, gain range, energy-resolution and $\alpha$-particle tracking response needed for the design of the next prototype stage. Operation at lower pressure, finer readout segmentation and in-beam validation are planned as subsequent developments. 

The $\alpha$-tracking performance of the Micromegas was compared to the results of numerical simulation, performed using the hydrodynamic approach~\cite{DasModel2025}, developed by us on \textsc{comsol} platform~\cite{Com}, to envisage the device dynamics and its suitability as a two-dimensional end-plate readout in the SAT-TPC.
The other numerical simulation work, relevant for the present work, were performed using \textsc{geant4}~\cite{Gea2003} and \textsc{garfield++}~\cite{Veenhof98}.

The article is organized in the following manner. Section~\ref{MM_proto} outlines the geometry of the bulk Micromegas opted for this work. The description of the experimental setup is furnished in Section~\ref{Exp_set}. The details of the characterization of the Micromegas by measuring the gas gain and energy resolution of the prototype for X-ray and $\alpha$-particles, carried out with the given filling gas mixtures, are provided in Section~\ref{MM_char}. The performance of the prototype in measuring the $\alpha$-particle track is also described in this context. The section contains discussions in reference to the numerical simulations, as used for determination of the gas gain and tracking performance. Section~\ref{Con} concludes with a summary of the main findings and future projections.

\section{Micromegas Prototype}
\label{MM_proto}
\begin{figure}[h!]
    \centering
    \begin{subfigure}[b]{0.4\textwidth}
        \centering
        \includegraphics[height=5cm,width=\textwidth]{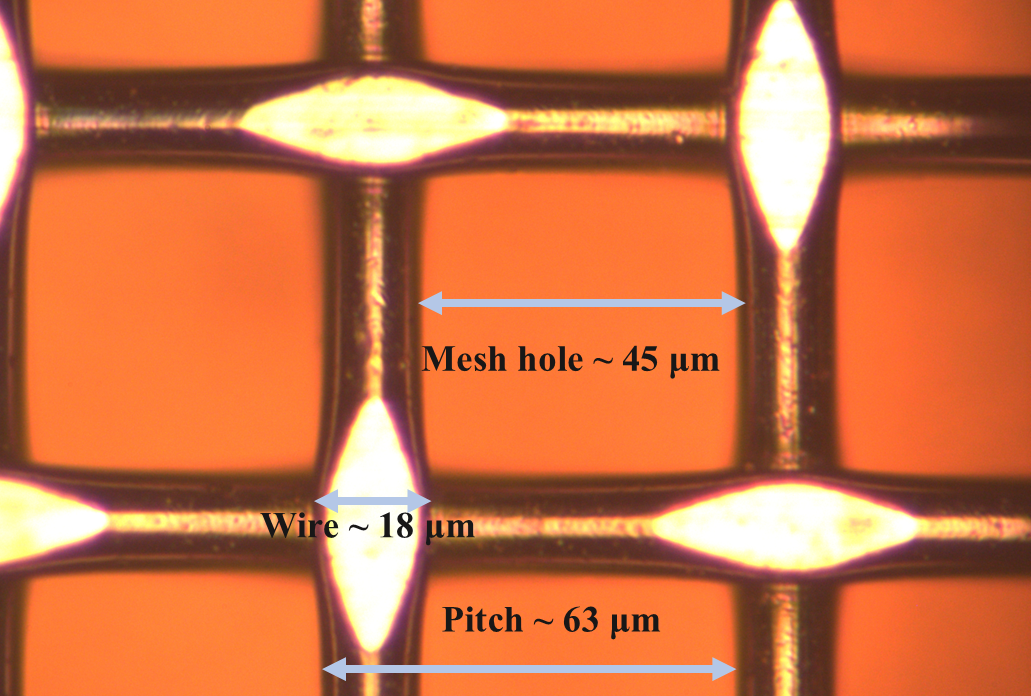}
        \caption{}
    \end{subfigure}
    % \hfill
    \begin{subfigure}[b]{0.35\textwidth}
        \centering
        \includegraphics[height=5cm,width=\textwidth]{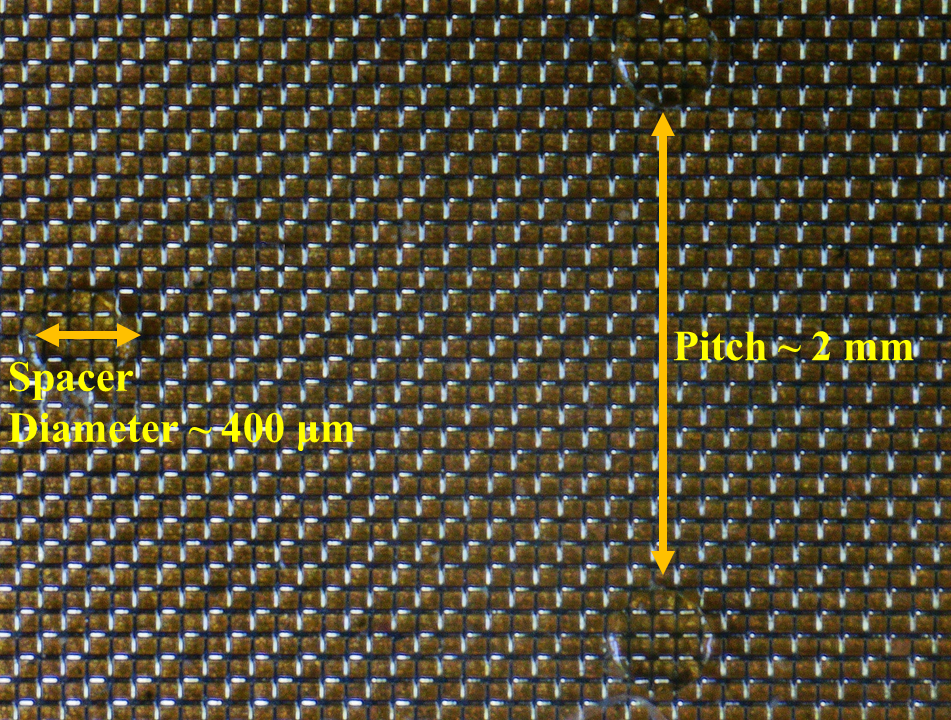}
        \caption{}
    \end{subfigure}
    \caption{Images of (a) micro-mesh made with stainless steel wires of 18~$\mu$m diameter and 63~$\mu$m pitch, and (b) supporting pillars of 400~$\mu$m diameter and 2~mm pitch}
    \label{fig:MM_photos}
\end{figure}

A bulk Micromegas~\cite{Gio06}, where a micro-mesh is laminated directly onto a printed circuit board with amplification gap of 128~$\mu$m and active area $10 \times 10$~cm$^2$ was selected as the end-plate readout to be integrated in the prototype SAT-TPC setup. In the Micromegas, the micro-mesh separates the drift region from the amplification region. The images of the Micromegas prototype, taken with a calibrated microscope \textit{Olympus MX51}~\cite{OlympusMX51} using optical zoom 50X and 10X, respectively, are shown in figure~\ref{fig:MM_photos}. It shows that the micro-mesh of the Micromegas is made of calendered stainless steel wires of diameter 18~$\mu$m with 63~$\mu$m pitch. The spacers, used to maintain the amplification gap and mechanical rigidity of the mesh, are of diameter 400~$\mu$m and fixed with a periodicity 2~mm. The anode plane consists of ten readout strips, each 9.5~mm wide, separated by 0.5~mm gaps, corresponding to a 10 mm pitch.

\section{Experimental Setup}
\label{Exp_set}
\begin{figure}[h!]
    \centering
    \includegraphics[height= 7cm ,width=0.8\textwidth]{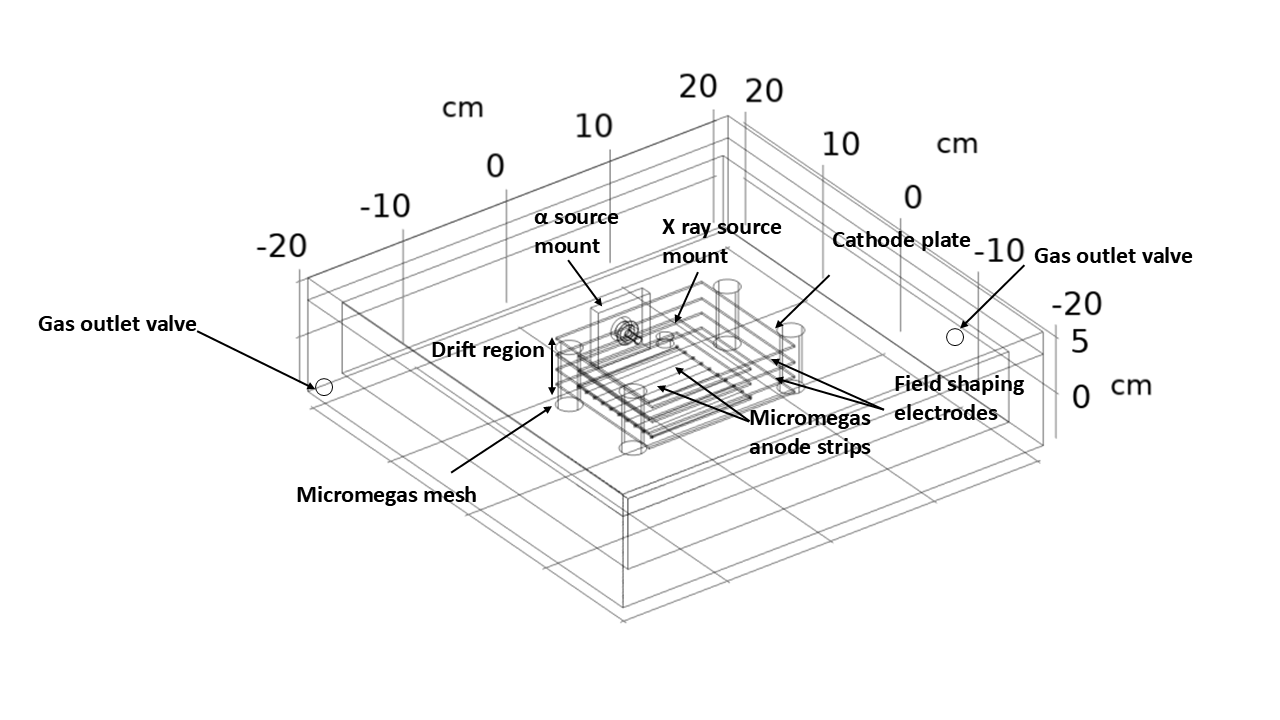}
    \caption{Schematic diagram of the experimental setup}
    \label{fig:SAT-TPC_schematic}
\end{figure}

\begin{figure}[h!]
    \centering
    \begin{subfigure}[b]{0.45\linewidth}
        \centering
        \includegraphics[height=5cm,width=\textwidth]{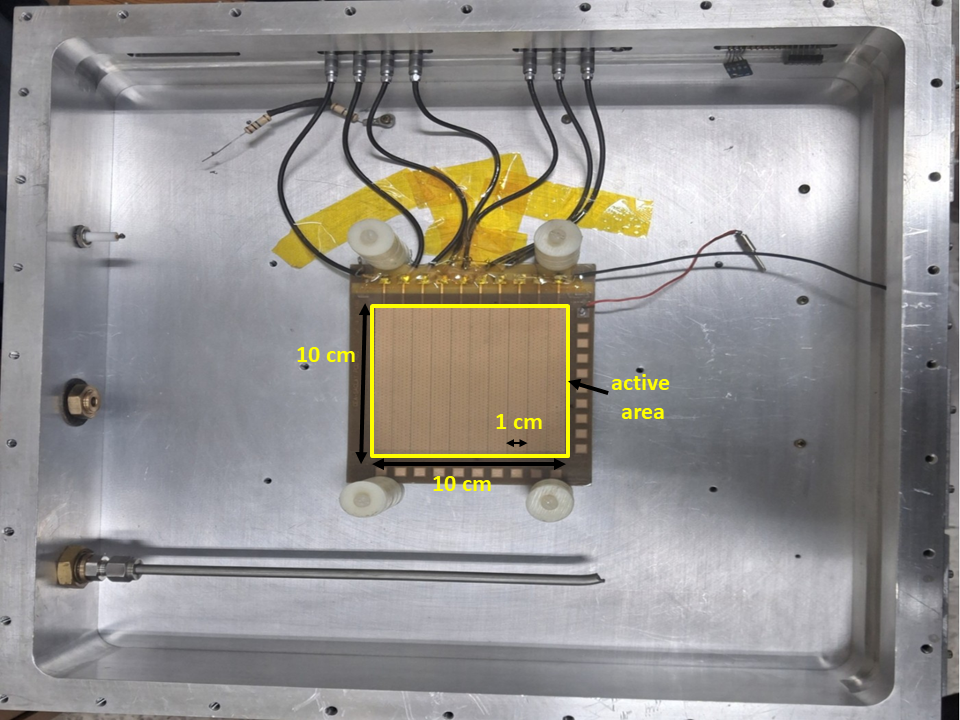}
        \caption{}
        \label{p_SAT-TPC_xray}
    \end{subfigure}
    \begin{subfigure}[b]{0.45\linewidth}
        \centering
        \includegraphics[height=5cm,width=\textwidth]{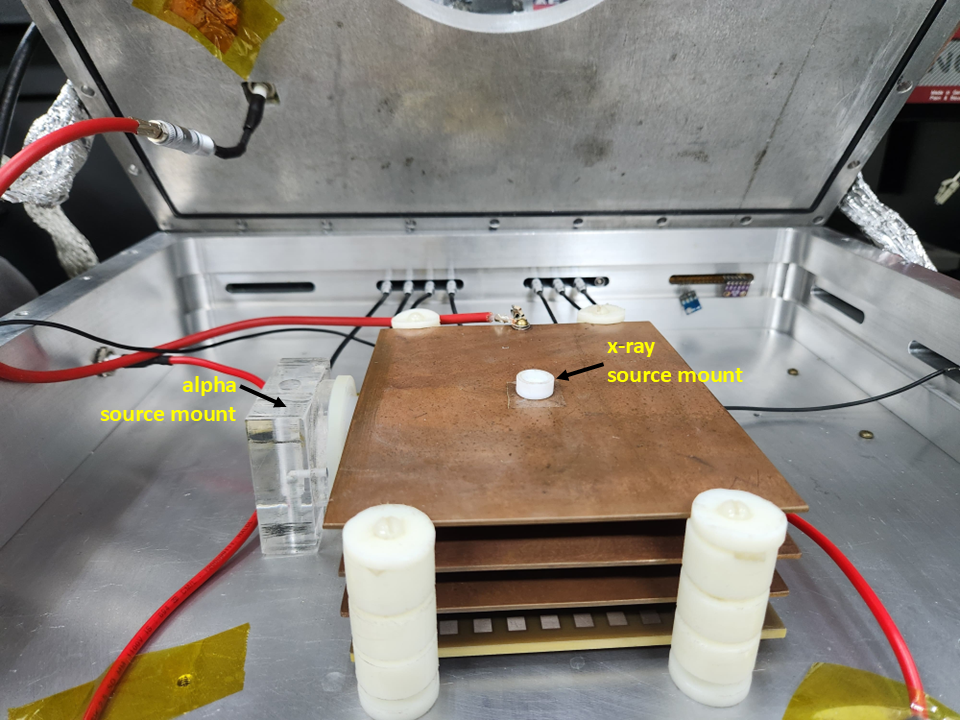}
        \caption{}
        \label{events_xray}
    \end{subfigure}
    \caption{(a) Micromegas prototype, and (b) the experimental setup with field cage, X-ray and $\alpha$-particle source}
    \label{p_SAT-TPC_figure}
\end{figure}

%The characterized Micromegas was then integrated into a SAT-TPC prototype with a 4~cm drift region. This configuration allowed the readout response to be studied in the presence of an extended drift volume, field-shaping electrodes and gas circulation. 
The Micromegas prototype was installed in a chamber with internal volume $6.5 \times 40 \times 40$~cm$^3$, equipped with a continuous gas flow rate of 0.5L/hour of the pre-mixed Ar + CO$_2$ (90:10) and Ar + i-C$_4$H$_{10}$ (95:5). To configure an extended drift volume, a copper plate was positioned 4~cm above the Micromegas as the drift cathode. Two field-shaping electrodes, machined from copper plate, were placed 1.34~cm apart between the cathode and the micro-mesh of the Micromegas to produce a field cage in the drift volume and facilitate maintaining a uniform electric field throughout. 
The drift region was irradiated with 5.9~keV X-rays from a $^{55}$Fe source and with 5.48~MeV $\alpha$-particles from a $^{241}$Am source. The $^{55}$Fe X-ray source was placed on the cathode plate with a hole at its center to irradiate the active volume in longitudinal direction. A holder made from perspex was built for mounting the $^{241}$Am $\alpha$-source. It was positioned on one side of the setup and outside the field-shaping electrodes, holding the source 2~cm above the mesh plane to project the $\alpha$-particles in the transverse plane of the drift volume of the prototype to produce the tracks in orthogonal direction to the readout strips on the anode plane. The source was collimated by a nylon tube of 2~mm diameter. 
The physical setup is schematically represented in figure~\ref{fig:SAT-TPC_schematic} with all the components labeled.
The pictures of the Micromegas and the experimental setup are shown in figure~\ref{p_SAT-TPC_figure} where the location of the radioactive sources are marked. 

\begin{figure}[h!]
        \centering
        \includegraphics[width=0.5\textwidth]{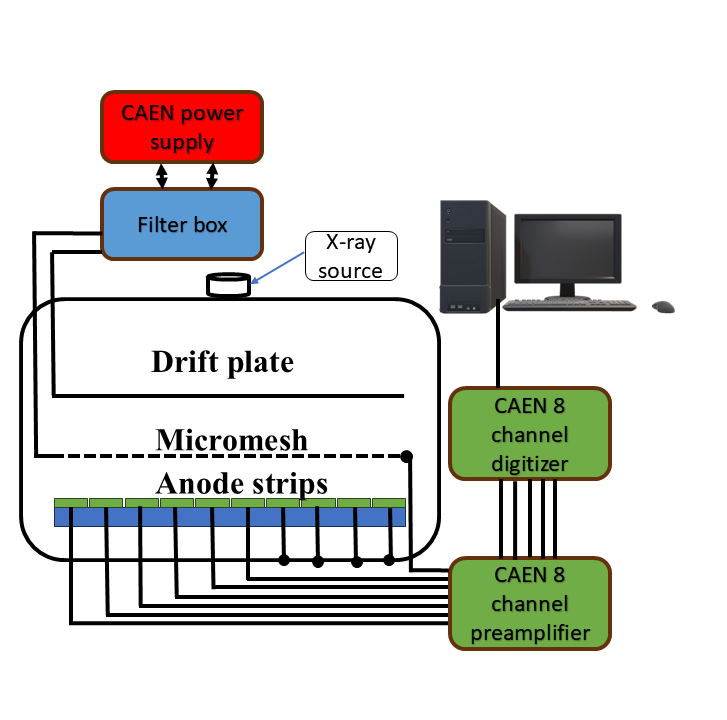}
    \caption{Scheme of the experimental setup}
    \label{setup}
\end{figure}

The drift cathode and the two field-shaping electrodes were connected together through a voltage divider utilizing 10 M$\Omega$ resistors.
%The same electronic setup, as discussed in sub-section~\ref{test_set}, was implemented here as well. 
The drift electric field was configured by applying negative high voltage to the drift cathode plate using power supply \textsc{CAEN DT5521E}~\cite{CAENDT5521E}, while the micro-mesh was also negatively biased with respect to the grounded anode readout to configure the amplification field. In the anode plane, six readout strips out of ten were used for data collection due to limited number of readout electronics, though all ten strips were kept grounded. The six readout strips and the micro-mesh were connected to a low-noise 8-channel charge-sensitive preamplifier \textsc{CAEN A1422}~\cite{CAENA1422} followed by a digitizer \textsc{CAEN V1730S}~\cite{CAENV1730S} for signal recording aided by \textsc{CoMPASS} software. In accordance with the Ramo-Shockley theorem, the movement of both electrons and ions within the amplification gap induces a signal on the mesh. This signal is characterized by a fast, prompt component induced by the electrons, followed by a long, delayed tail contributed by the ions. This integrated signal is approximately equal and opposite to the charge that would be induced on an unsegmented anode plane, and consequently, it was utilized as the trigger for data acquisition.
In this setup, the trigger logic was configured to record an event when both the micro-mesh and the readout channel fired within a 1~$\mu$s window. A schematic representation of the experimental setup is shown in figure~\ref{setup}.

\section{Micromegas Characterization}
\label{MM_char}
\subsection{Electron Transmission} 
To optimize the operating voltage configurations, measurement of electron transmission ($T$) of the Micromegas plays an important role. It can be defined as the fraction of primary electrons that successfully pass through the micro-mesh into the amplification gap from the drift region. Mathematically, it can be expressed as:
\begin{equation}
T = \frac{N_{amp}}{N_{drift}}
\end{equation}
where $N_{amp}$ is the number of electrons that reach the amplification region, while $N_{drift}$ is the total number of primary electrons arriving at the micro-mesh from the drift region. 

%A systematic investigation of this parameter was carried out by varying the drift field ($E_{drift}$) or the amplification field ($E_{amp}$) while keeping the other one fixed. 
The electron transmission of the present 128~$\mu$m bulk Micromegas in use was determined experimentally by measuring the anode signal amplitude as a function of drift field while keeping the amplification field fixed. Then, the change in the signal amplitude or the centroid of the $^{55}$Fe full-energy peak due to the variation in the drift electric field should be proportional to the number of the primary electrons successfully extracted through the micro-mesh. The observed variation in the peak centroids with the drift field was normalized by the largest centroid value, which represents the maximum transmission point. The amplification field was held constant at 40.8~kV/cm for Ar + CO$_{2}$ (90:10) and 26.5~kV/cm for Ar + i-C$_4$H$_{10}$ (95:5). These optimized values were pre-determined through initial gain measurements to provide a sufficient signal-to-noise ratio for the $^{55}$Fe-peak while ensuring stable detector operation well below the micro-discharge threshold for each of the respective gas mixtures. While keeping the amplification fields fixed, the drift field was systematically varied across a range of approximately 100 to 1000~V/cm in each case.

\begin{figure}[h!]
    \centering
    \includegraphics[width=0.5\textwidth]{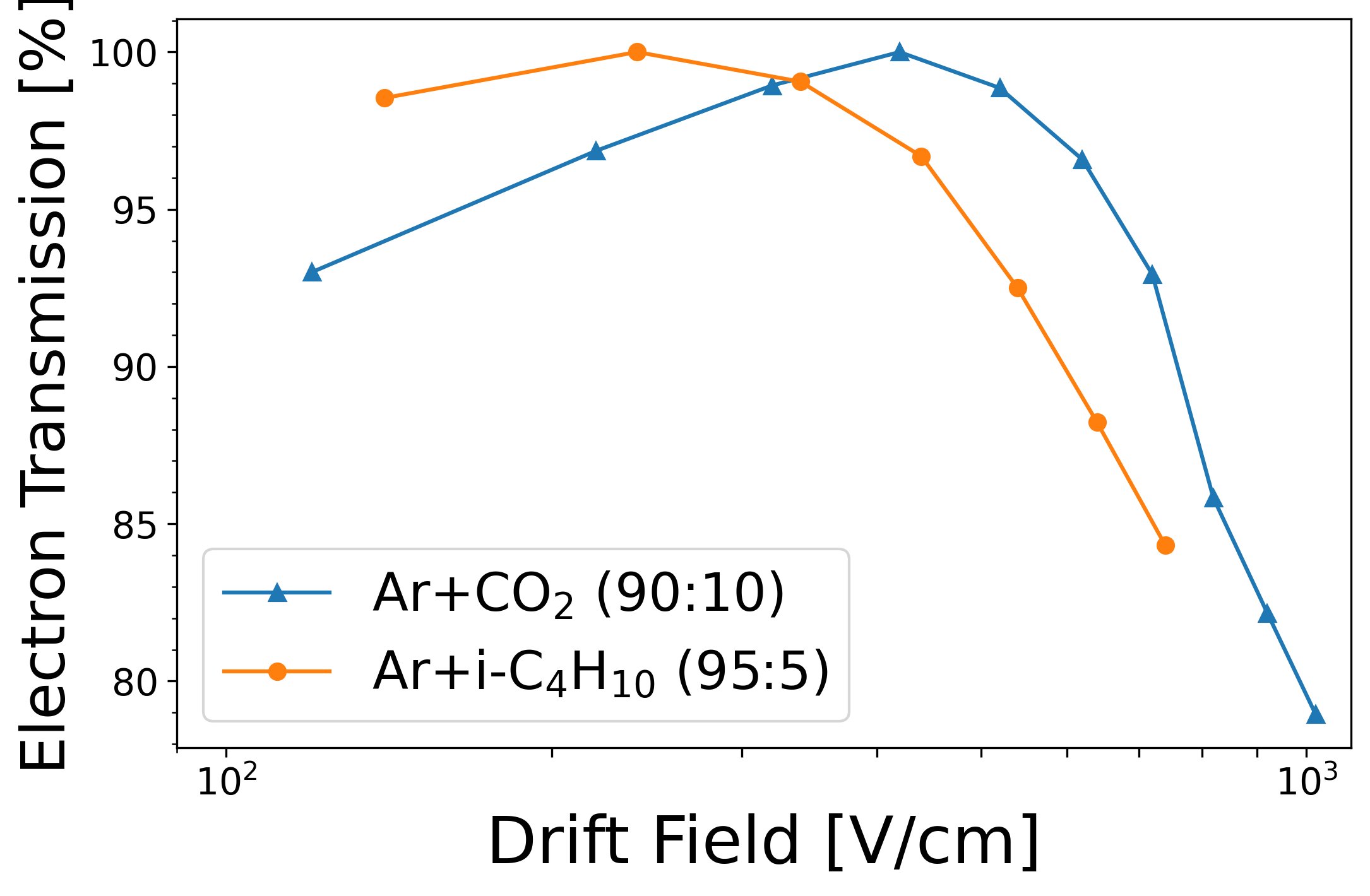}
    \caption{Electron transmission as a function of drift field for the 128~$\mu$m bulk Micromegas, performed with amplification field 40.8~kV/cm in Ar + CO$_2$ (90:10) and 26.5~kV/cm in Ar + i-C$_4$H$_{10}$ (95:5)}
    \label{fig:transparency_plot}
\end{figure}

The normalized electron transmission values, for the two gas mixtures are plotted in figure~\ref{fig:transparency_plot} as a function of the drift field. The normalized electron transmission increases with the drift field up to a maximum-transmission region and subsequently decreases at higher drift fields. The comparison with the previous work~\cite{Bhattacharya2014} was made because 
electron transmission was studied for a bulk Micromegas with the same two gas 
mixtures, providing a relevant reference for the present field-ratio measurements. The results show similarity to those reported in~\cite{Bhattacharya2014}. At very low drift fields, the normalized transmission is lower because some electrons may be lost through recombination or attachment to trace electronegative impurities before reaching the mesh. As summarized in Table 1, the optimal drift fields for the applied amplification fields, yielding the field ratios ($E_{drift}/E_{amp}$) required to reach the maximum were determined to be 0.0098 for Ar + CO$_{2}$ (90:10) and 0.0094 for Ar + i-C$_4$H$_{10}$ (95:5). At higher drift fields, the transparency declines as the electric field lines from the drift region increasingly terminate directly on the metallic wires of the micro-mesh rather than passing through the holes. As a result, the primary electrons following the field lines collide with the mesh and are absorbed before entering the amplification volume. The region around the maximum normalized transmission in figure~\ref{fig:transparency_plot} defined the optimal operating regime for the present Micromegas. This region was selected as the operating condition because it maximizes the relative electron transmission through the mesh. Consequently, an average field ratio of $\approx 9.6 \times 10^{-3}$, which closely approximates these experimental optima, was adopted as the reference configuration for the subsequent measurement.
\begin{table}[h!]
\centering
\caption{Optimal field ratio for maximum electron transparency as measured in this work and in ~\cite{Bhattacharya2014}}
\label{tab:field_ratios}
\begin{tabular}{|l|c|c|}
\hline
 & Ar + CO$_{2}$ (90:10) & Ar + i-C$_{4}$H$_{10}$ (95:5) \\ \hline
Previous Work~\cite{Bhattacharya2014} & 0.012 & 0.011 \\ \hline
This Work & \textbf{0.0098} & \textbf{0.0094} \\ \hline
\end{tabular}
\end{table}

\subsection{Gas Gain with X-ray}
\label{gg_fe}
The prototype Micromegas setup was characterized by measuring its gas gain and energy resolution using the 5.9~keV X-ray from the $^{55}$Fe-source. The gas gain \(G\) can be defined as the ratio of the number of electrons collected to the number of primary electrons produced in the gas:
\begin{equation}
\label{GG}
    G = \frac{Q_{col}}{e \cdot N_p}
\end{equation}
where $Q_{col}$ is the collected charge, $e$ is the electronic charge and $N_p$ is the number of primary electrons. The number of the primary electrons, $N_p$, here was estimated using numerical simulation while the collected charge $Q_{col}$, was determined from the experimental measurements. 
During data collection, the detector chamber was maintained at a temperature of 24~$\pm$~1$^\circ$C and a pressure of 754~$\pm$~4.5 Torr.

\begin{figure}[htbp]
    \centering
    \begin{subfigure}[b]{0.45\textwidth}
        \centering
        \includegraphics[width=\textwidth]{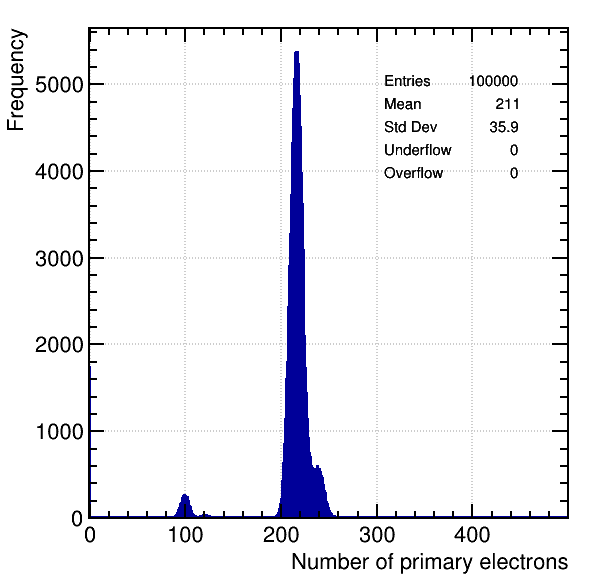}
        \caption{Ar + CO$_2$ (90:10)}
        \label{fig:fe55_arco2}
    \end{subfigure}
    \hfill
    \begin{subfigure}[b]{0.45\textwidth}
        \centering
        \includegraphics[width=\textwidth]{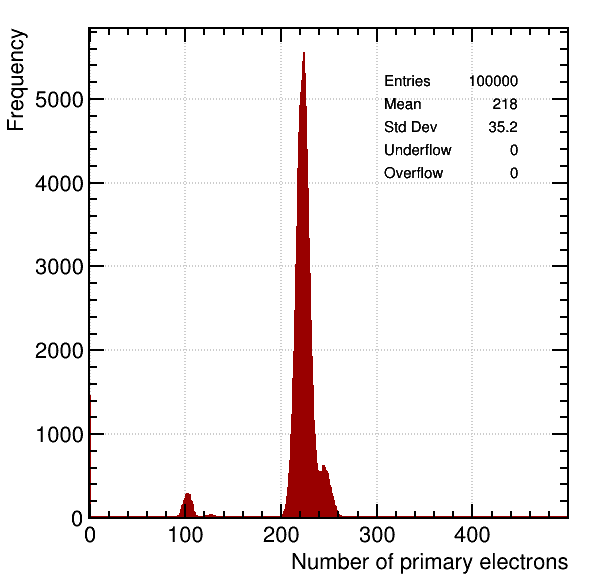}
        \caption{Ar + i-C$_4$H$_{10}$ (95:5)}
        \label{fig:fe55_aric4h10}
    \end{subfigure}
    \caption{Number of primary electrons produced by 5.9~keV X-rays in gas mixtures (a) Ar + CO$_2$ (90:10), and (b) Ar + i-C$_4$H$_{10}$ (95:5)}
    \label{fig:fe55_comparison}
\end{figure}

The number of the primary electrons, $N_p$, was estimated from \textsc{garfield++} code using the \textsc{heed} toolkit~\cite{Smirnov2005}, that implements \textit{Photo Absorption Ionization (PAI)} model. For each X-ray event, \textsc{heed} calculated the energy loss and generated ionization clusters along the photon interaction track. Each cluster corresponded to a localized group of electron-ion pairs created in a single ionization act, and the total number of primary electrons was obtained by summing the electrons from all clusters produced in an event.
The histograms of primary ionization electrons produced by the 5.9~keV X-ray in Ar + CO$_2$ (90:10) and Ar + i-C$_4$H$_{10}$ (95:5) gas mixtures are shown in figure~\ref{fig:fe55_comparison}, as obtained from \textsc{garfield++}. The resulting distributions exhibited mean values of 211 and 218 primary electrons, respectively, which were considered as $N_{p}$ values for the corresponding gas mixtures. The simulated distributions accurately capture the underlying physics of the X-ray interactions. Notably, the small secondary peak at $\sim$100 electrons corresponds to the Argon escape peak, occurring when the $\sim$2.9~keV Argon fluorescence photon escapes the active volume. Furthermore, the high-energy shoulder on the main peak is caused by the $\approx$12\% contribution of the 6.49~keV $K_{\beta}$ emission line from the $^{55}Fe$ source.

\begin{figure}[h!]
     \centering
     \begin{subfigure}[b]{0.45\linewidth}
         \centering
         \includegraphics[height=5.5cm,width=\textwidth]{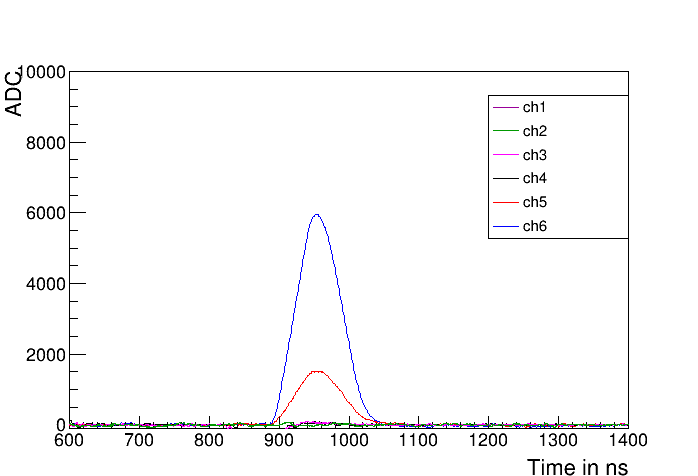}
         \caption{}
         \label{xray_strip}
     \end{subfigure}
     \begin{subfigure}[b]{0.45\linewidth}
         \centering
         \includegraphics[height=5cm,width=\textwidth]{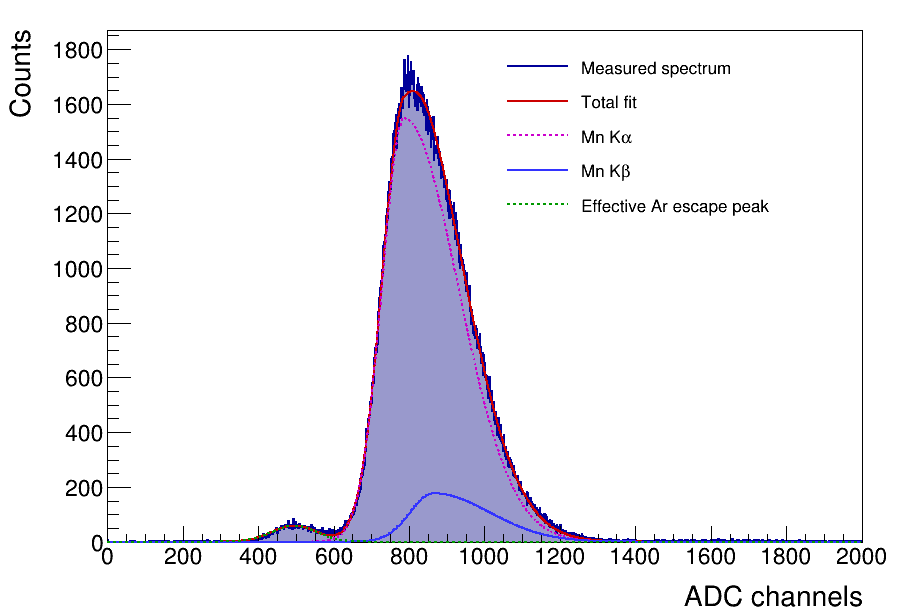}
         \caption{}
         \label{xray_spectrum}
     \end{subfigure}
   \caption{(a) Readout strip signals, and (b) energy spectrum of the $^{55}$Fe X-ray source recorded from the micro-mesh}
   \label{fig2}
\end{figure}

As an example of the detector's response when operated at an amplification field of 38.3~kV/cm for Ar + CO$_{2}$ (90:10) and 30.5~kV/cm for Ar + i-C$_{4}$H$_{10}$ (95:5), the fifth and sixth strips typically recorded the largest signals due to the location of the source, placed at the center, as shown in figure~\ref{xray_strip}. The corresponding energy spectrum was obtained from the micro-mesh signal recorded in the digitizer \textsc{CAEN V1730S} following pre-amplification with \textsc{A1422}, as shown in figure~\ref{xray_spectrum}.
The charge signals induced on the readout strips due to an X-ray event were converted into voltage and then amplified using 8-channel preamplifier \textsc{CAEN A1422}. The resulting analog voltages were digitized and recorded in the digitizer \textsc{CAEN V1730S} for offline analysis. 

\begin{figure}[h!]
\centering
\includegraphics[width=0.5\textwidth]{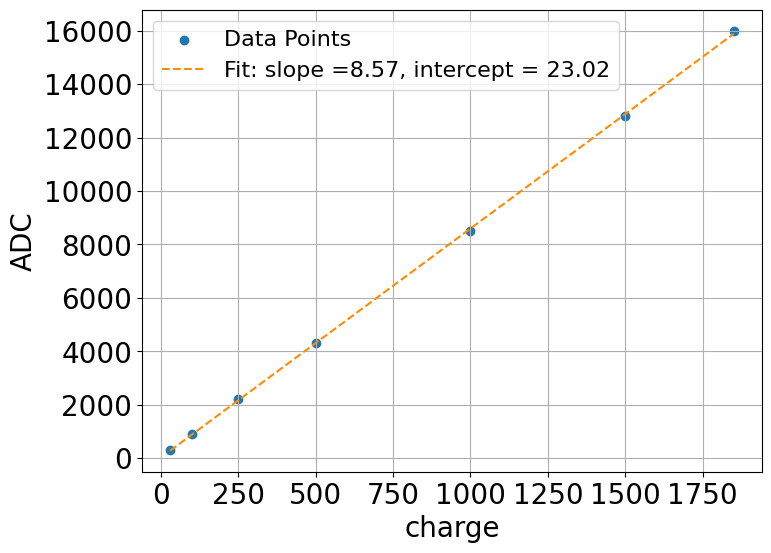}
\caption{Calibration plot used to convert ADC counts to collected charge}
\label{calib_track}
\end{figure}

During the offline analysis, the digitized waveforms corresponding to each event were integrated to extract the total ADC counts from which, collected charge, $Q_{col}$, was determined. For this purpose, an RC-injection calibration was performed through the entire electronic chain of the readout strips. The RC-injection calibration was performed on the discrete electronics chain used for processing the Micromegas signal, not connecting the detector in the electronics chain. Because the detector was not connected during the RC-injection
calibration, the detector capacitance and the capacitive coupling
between the readout strips and the micro-mesh were not included.
When the detector is connected, these additional capacitances can
modify the transient response of the preamplifier and may therefore
introduce a systematic uncertainty in the charge obtained from the
calibration and, consequently, in the absolute gas gain. The calibration plot of the charge versus ADC counts is shown in figure~\ref{calib_track}. The calibration fit, as shown in figure~\ref{calib_track}, was used to convert the ADC count to the injected charge using the following expression. 
\begin{equation}
\mathrm{charge\ [fC]} = \frac{\mathrm{ADC\ [counts]} - 23.02}{8.57} \,.
\label{charge_adc}
\end{equation} 

\begin{figure}[h!]
\centering
\includegraphics[height=6cm,width=0.7\textwidth]{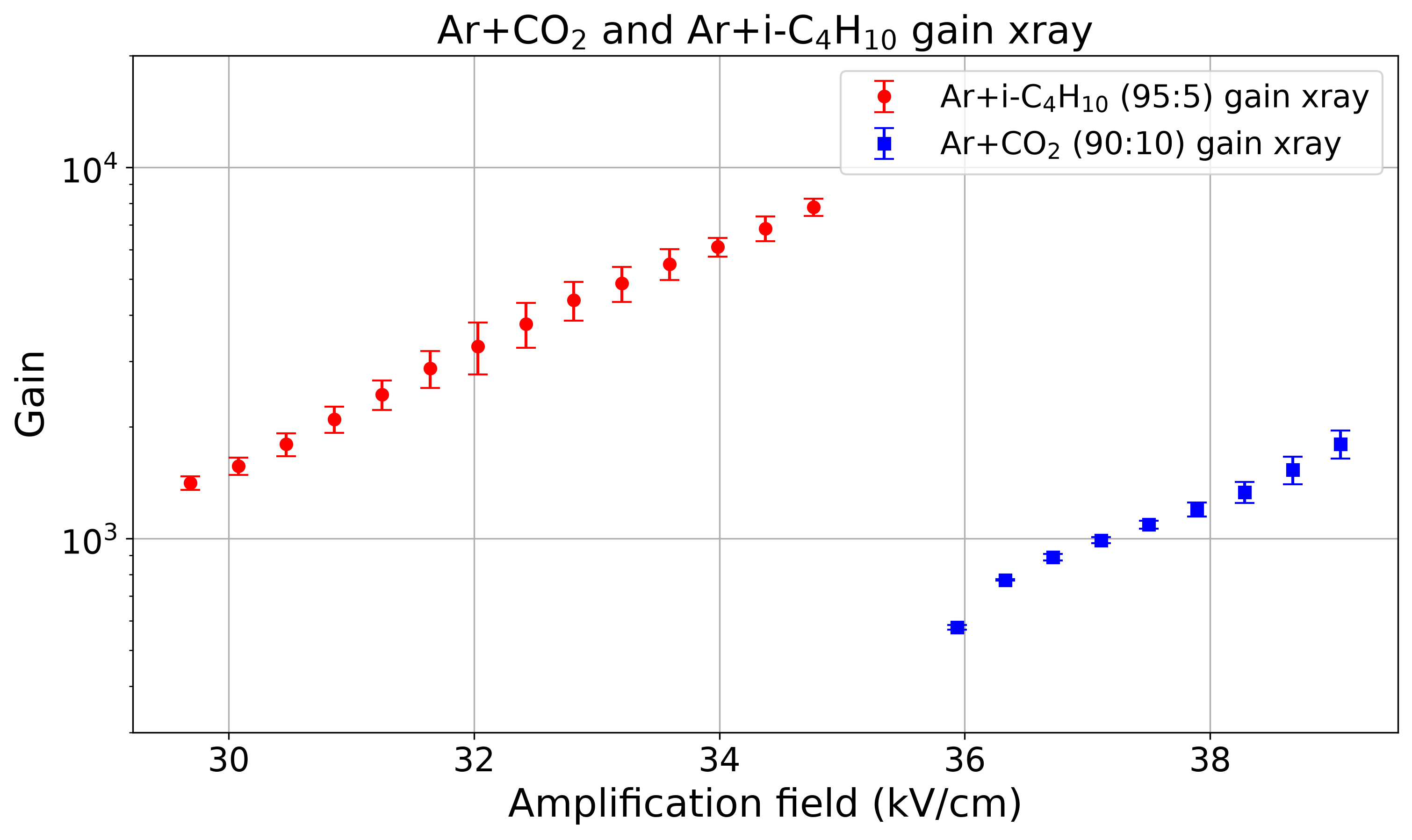}
\caption{Gain measured in Ar + CO$_2$ (90:10) and Ar + i-C$_4$H$_{10}$ (95:5) using the $^{55}$Fe X-ray source}
\label{gain_xray}
\end{figure}

The gas gain was subsequently evaluated by substituting the experimentally measured collected charge, $Q_{col}$, and the simulated average number of primary electrons, $N_{p}$, into equation~\ref{GG}.
To study the variation in Micromegas gain with the field configuration, both the cathode and mesh voltages were varied systematically while maintaining equal potential differences between field-shaping electrodes and the micro-mesh. The gain was determined as a function of the amplification field by systematically varying the micro-mesh voltage from approximately -460~V to -500~V for Ar + CO$_{2}$ (90:10), and from -380~V to -445~V for Ar + i-C$_{4}$H$_{10}$ (95:5). The corresponding cathode voltages were adjusted simultaneously to maintain the optimal field ratio. 
Figure~\ref{gain_xray} presents the gain curves for Ar + CO$_2$ (90:10) and Ar + i-C$_4$H$_{10}$ (95:5), measured with 5.9~keV X-rays from $^{55}$Fe-source where the results are comparable to the previous work reported in~\cite{Bhattacharya2014}. Both mixtures exhibited the expected exponential dependence of gain on the amplification field. For Ar + CO$_2$, noticeable amplification started at approximately 36~kV/cm, whereas Ar + i-C$_4$H$_{10}$ (95:5) showed similar gain at comparatively lower field value and offered comparatively higher gain. The higher gain observed with the Ar + i-C$_4$H$_{10}$ (95:5) mixture is attributed to the enhanced Penning effect at elevated field values, arising from the lower ionization potential of the isobutane molecules relative to the metastable states of argon.

\subsection{Energy Resolution with X-ray}
The $^{55}\mathrm{Fe}$ spectrum contains unresolved Mn K$\alpha$ and K$\beta$ full-energy contributions, together with a lower-energy feature associated with argon escape events. The Mn K$\alpha$ and K$\beta$ contributions were described using asymmetric Gaussian functions to account for the observed asymmetry of the full-energy peak, while the lower-energy feature was represented by an effective Gaussian component. The integrated K$\beta$/K$\alpha$ area ratio was constrained to approximately 12\%, consistent with the HEED simulation shown in figure~\ref{fig:fe55_comparison}. The total fitted response, obtained by summing the three components, is shown in figure~\ref{xray_spectrum}. The energy resolution was extracted from Gaussian fits to the $^{55}$Fe full-energy peak and is quoted as $\sigma/\mu$, where $\mu$ and $\sigma$ are the fitted peak mean and standard deviation. The values obtained were $9.7\%$ for Ar + CO$_2$ and $8.5\%$ for Ar + i-C$_{4}$H$_{10}$, corresponding to FWHM resolutions of about $22.9\%$ and $19.9\%$, respectively. The better resolution observed with the latter mixture reflects improved statistical charge collection and reduced gain fluctuations, consistent with the trends reported in GEM-based studies \cite{Roy2021} and Micromegas performance measurements \cite{Bhattacharya2014}.

\subsection{$\alpha$-Particle Tracking}
\begin{figure}[htbp]
    \centering
    \begin{subfigure}[t]{0.48\linewidth}
        \centering
        \includegraphics[height=5cm]{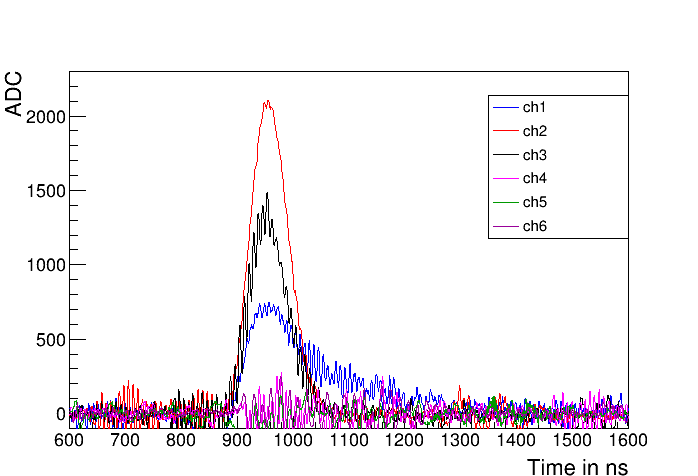}
        \caption{}
        \label{pulse_example}
    \end{subfigure}
    \hfill
    \begin{subfigure}[t]{0.48\linewidth}
    \centering
    \includegraphics[height=5cm]{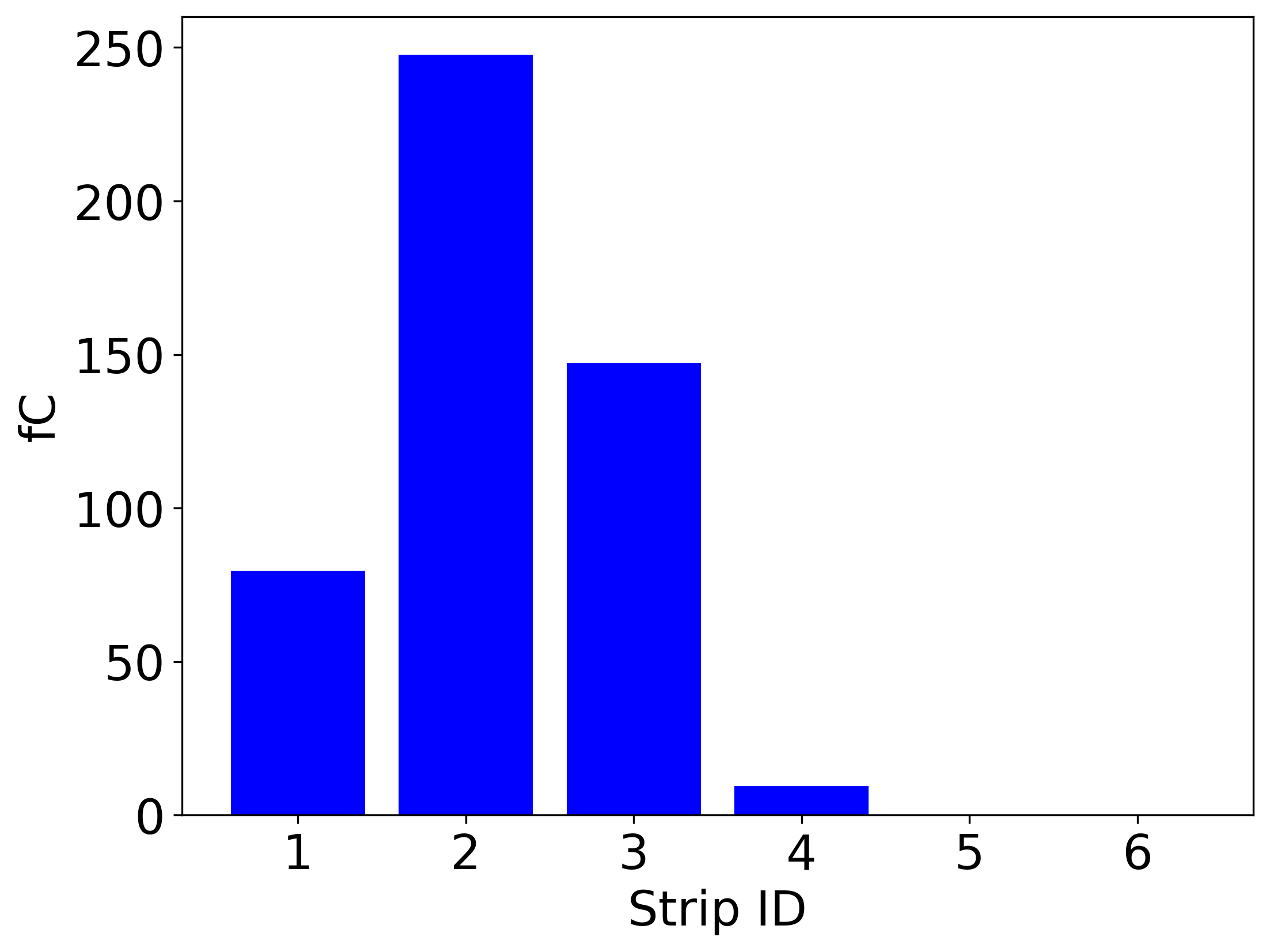}
    \caption{}
    \label{pulse_profile}
\end{subfigure}
\caption{(a) ADC counts from six readout strips, and (b) charge collected on the individual readout strips in fC for an $\alpha$-event, measured with Ar + CO$_2$ (90:10) gas mixture at amplification field 28.5~kV/cm.} 
\label{pulse}
\end{figure}

The prototype setup was irradiated with 5.48~MeV $\alpha$-particles from the $^{241}$Am-source in perpendicular direction with respect to the readout strips to study the Micromegas response using both the gas mixtures as shown in figure \ref{fig:SAT-TPC_schematic}. 
A typical set of signals obtained experimentally from individual six readout strips of the Micromegas is shown in figure~\ref{pulse_example} when operated with the Ar + CO$_2$ (90:10) gas mixture with amplification field of 28.5~kV/cm. The waveforms display a distinct variation in pulse height across the channels. This variation is primarily driven by the specific energy loss profile (Bragg curve) of the $\alpha$-particle as it traverses the gas volume, depositing the highest concentration of charge near the end of its track. Additionally, transverse diffusion of the drifting electron cloud causes charge sharing across the boundaries of neighboring strips, which further modulates the final collected charge on each channel. The corresponding integrated ADC counts, as shown in figure ~\ref{pulse_profile}, provide the charge-deposit profile on the readout plane, a distribution that is subsequently corroborated by the numerical simulations.

\begin{figure}[htbp]
    \centering
    \begin{subfigure}[b]{0.32\linewidth}
        \centering
        \includegraphics[width=\textwidth]{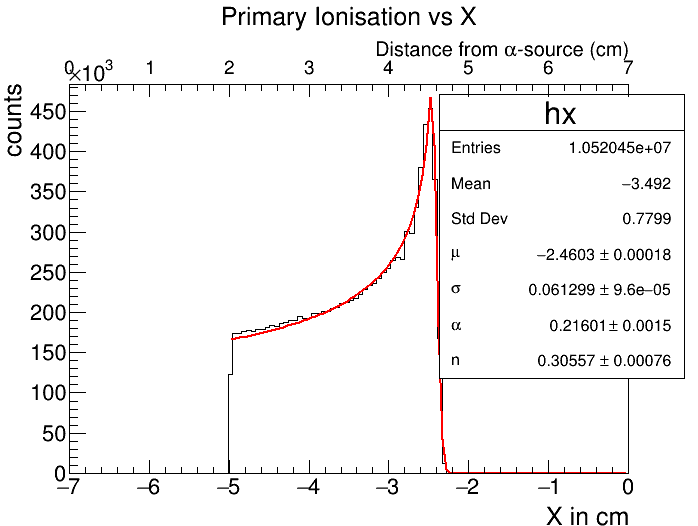}
        \caption{Longitudinal (X)}
        \label{fig:geant_x}
    \end{subfigure}
    \hfill
    \begin{subfigure}[b]{0.32\linewidth}
        \centering
        \includegraphics[width=\textwidth]{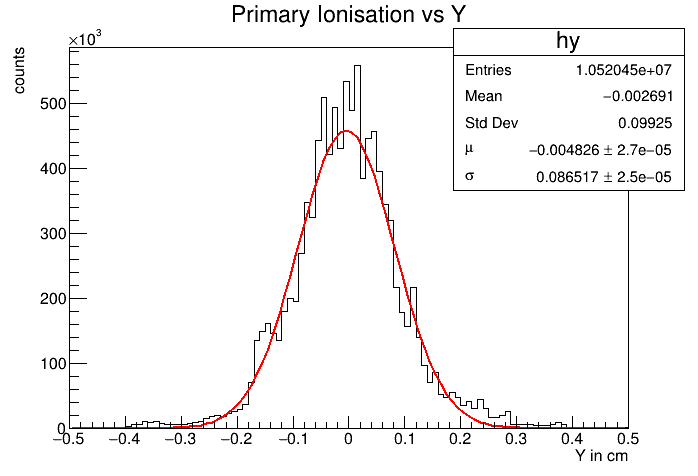} % Update filename if needed
        \caption{Transverse (Y)}
        \label{fig:geant_y}
    \end{subfigure}
    \hfill
    \begin{subfigure}[b]{0.32\linewidth}
        \centering
        \includegraphics[width=\textwidth]{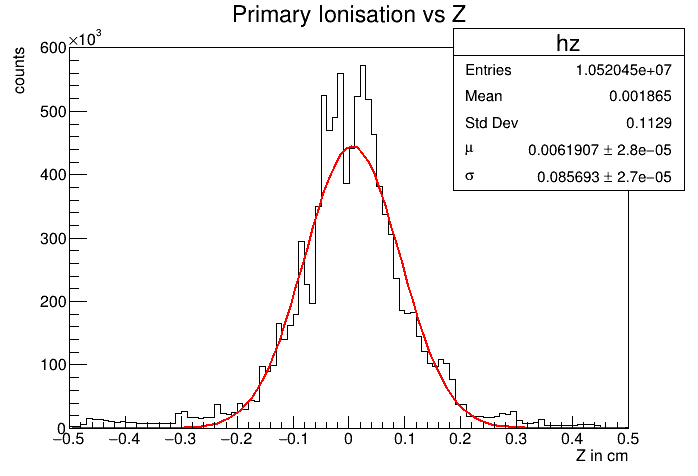} % Update filename if needed
        \caption{Transverse (Z)}
        \label{fig:geant_z}
    \end{subfigure}
    \caption{Spatial distributions of the primary ionization produced by a 5.48 MeV $\alpha$-particle simulated using \textsc{Geant4} along the (a) longitudinal $X$, (b) transverse $Y$, and (c) transverse $Z$ directions. In (a), the lower axis represents the detector-centered $X$ coordinate, while the upper axis gives the distance from the $\alpha$-source, with the source position defined as zero.} 
    \label{sim_geant}
\end{figure}

To follow the device dynamics, the measured profile of the charge deposit by the 5.48~MeV $\alpha$-particle was simulated using the hydrodynamic model, used to optimize and study the performance of the SAT-TPC prototype~\cite{DasModel2025}. The primary ionization caused by 5.48~MeV $\alpha$-particles from the $^{241}$Am-source was simulated in \textsc{geant4} using low-energy physics lists \textit{(Penelope, Livermore, PAI)}, considering the exact geometry of the source position in the experimental setup. The range of the $\alpha$-particle in both gases is approximately 5~cm, and accounting for the energy lost while traversing the 2~cm distance from the source location to the active volume, the effective energy deposited inside the experimental setup was calculated to be about 3.43~MeV.
To fully characterize the spatial extent of this initial charge deposition, the simulated primary ionization was mapped in three dimensions. Figure~\ref{sim_geant} illustrates the spatial distribution of the primary electrons along the longitudinal (X) and transverse (Y and Z) axes within the active volume. The longitudinal X-distribution captures the energy loss profile, terminating in the characteristic Bragg peak, while the Y and Z distributions demonstrate the tight initial transverse confinement of the $\alpha$-particle track before any diffusion occurs.
To generate a macroscopic seed cluster of electronic charges for the subsequent transport simulation, this X-spatial distribution was parameterized and fitted (marked by the red line in figure~\ref{sim_geant}) using a Crystal Ball function. This specific function was selected because it accurately models the asymmetric energy-loss distribution characteristic of heavy charged particles, seamlessly combining a Gaussian core for the Bragg peak with a power-law tail to account for energy straggling. The empirical form of the function is defined as:
\begin{equation}
f(x; \alpha, n, \bar{x}, \sigma) = \begin{cases} 
\exp\left(-\frac{(x-\bar{x})^2}{2\sigma^2}\right), & \text{for } \frac{x-\bar{x}}{\sigma} > -\alpha \\
A \cdot \left(B - \frac{x-\bar{x}}{\sigma}\right)^{-n}, & \text{for } \frac{x-\bar{x}}{\sigma} \le -\alpha 
\end{cases}
\label{eq:crystal_ball}
\end{equation}
\noindent where $\bar{x}$ and $\sigma$ denote the mean position and the standard deviation of the Gaussian core, respectively. The parameters $\alpha$ and $n$ govern the transition threshold to the power-law tail and its steepness. To guarantee the continuity of both the function and its first derivative at the transition boundary, the normalization constants $A$ and $B$ are constrained as follows:
\begin{equation}
A = \left(\frac{n}{|\alpha|}\right)^n \exp\left(-\frac{|\alpha|^2}{2}\right), \quad B = \frac{n}{|\alpha|} - |\alpha|.
\label{eq:crystal_ball_const}
\end{equation}
The other transverse spatial distributions were fitted with Gaussian.

The propagation of the parameterized three-dimensional electronic charge cluster through the drift volume was then simulated using a hydrodynamic approach. Implemented in the \textsc{comsol} platform using the \textit{Transport of Dilute Species (TDS)} module, this model treats the primary charges (electrons or ions) as charged solutes in a dilute solvent of neutral gaseous molecules. The temporal evolution of the seed cluster is governed by the drift-diffusion-reaction equation:
\begin{equation}
  \frac{\partial c_{j}}{\partial t}
  + \nabla\!\cdot\!\big(-D_j\,\nabla c_j + \mathbf{u}_{j}\, c_j\big)
  = S_j,
  \label{eq:driftdiffusion}
\end{equation}
where $c_{j}$, $D_{j}$, $\mathbf{u}_{j}$, and $S_{j}$ denote the concentration, diffusion coefficient, drift velocity, and net rate of production (source/sink term), respectively, for the charged species $j$. This hydrodynamic approach enables efficient macroscopic modeling of the drift and diffusion processes of the charge carriers. The necessary transport parameters ($D_{j}$ and $\mathbf{u}_{j}$) for the given gas mixtures were extracted from \textsc{magboltz}~\cite{Biagi1999}, based on the specific electric field configuration of the setup.
It should be noted that the woven geometry of the micro-mesh was not explicitly included in the macroscopic hydrodynamic model, since the primary objective of the simulation was to describe the macroscopic transport and spatial distribution of the primary electron cloud over the readout strips rather than the microscopic processes occurring within the Micromegas amplification region. The micro-mesh boundary was therefore treated as fully transparent to the drifting electrons, and electron multiplication within the amplification gap was not included. Consequently, the model does not account for electron losses at the mesh or avalanche fluctuations, which may affect the absolute collected charge.
%The entire 3D primary charge distribution was parameterized and fitted with a Crystal Ball function (marked by the red line) to generate a macroscopic seed cluster for the subsequent transport simulation \cite{DasModel2025}. The propagation of this charge cluster through the TPC was modeled using a hydrodynamic approach implemented in the \textsc{comsol} platform, which treats the electron cloud as solute particles in a solvent governed by drift-diffusion equations. The necessary macroscopic transport parameters, namely drift velocity and diffusion coefficients for the given gas mixtures, were extracted from \textsc{magboltz}~\cite{Biagi1999} based on the specific electric field configuration of the setup.

\begin{figure}[htbp]
\centering
\includegraphics[width=0.7\textwidth]{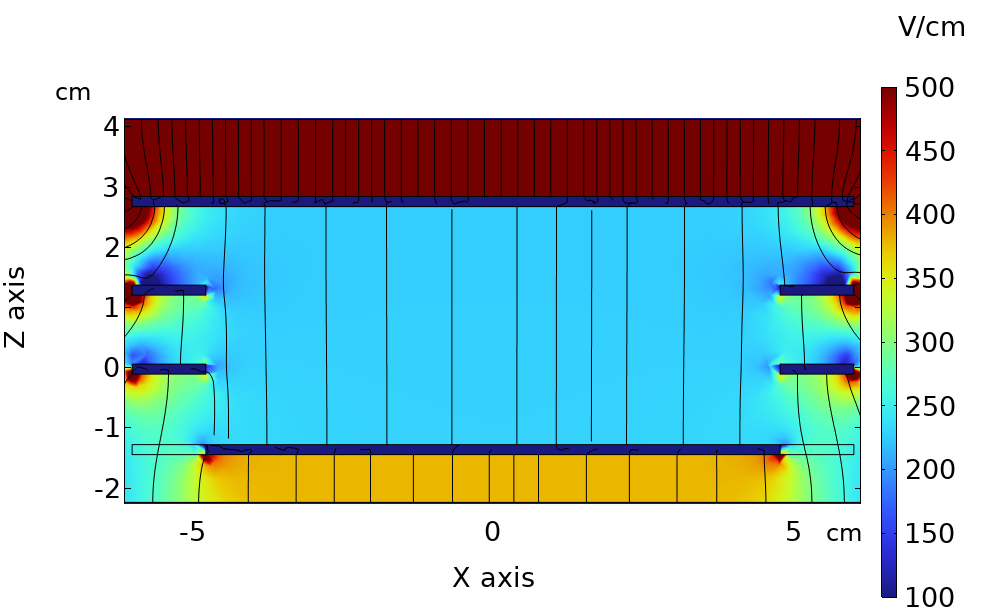}
\caption{Electric-field magnitude (in V/cm) on the longitudinal cross-section ($XZ$-plane) of the SAT-TPC prototype. The black curves represent magnitude-controlled electric-field streamlines.}
\label{field_map}
\end{figure}

The electric field map in the experimental setup was simulated using \textsc{comsol}. The result is shown in figure~\ref{field_map} where the electric field strength is represented by the color code, while the black lines depict the field lines. It may be observed that in the central drift region, the simulated electric field lines are straight and parallel, confirming a uniform drift field. The edge effects are visible near the edges and around the field-shaping electrodes indicating localized field distortions. For the applied voltages of -1200~V at the drift cathode and -300~V at the micro-mesh, the electric field in the central active region is approximately 228~V/cm.
The ratio of the drift to amplification field, used in the calculation, was as follows:
\[
\frac{E_{\mathrm{drift}}}{E_{\mathrm{amp}}} \approx 9.6 \times 10^{-3},
\]

\begin{figure}[htbp]
    \centering
    \begin{subfigure}[b]{0.42\linewidth}
        \centering
        \includegraphics[width=\textwidth]{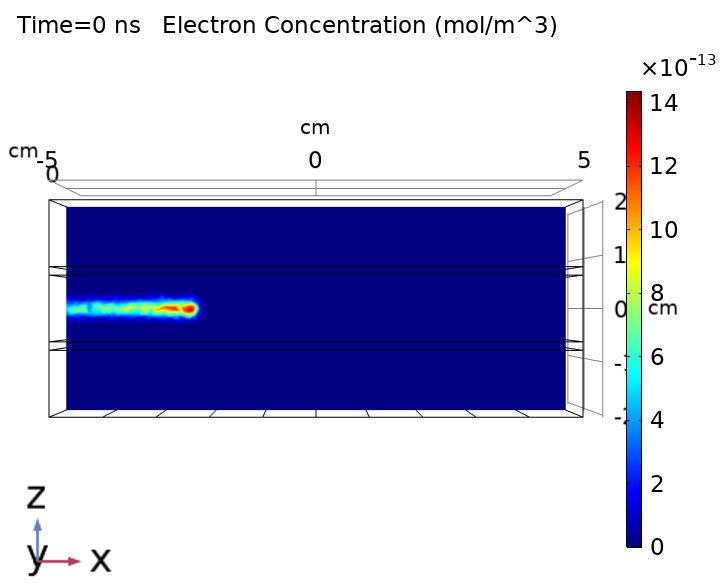}
        \caption{$t = 0$ ns}
        \label{fig:comsol_0}
    \end{subfigure}
    \hfill
    \begin{subfigure}[b]{0.48\linewidth}
        \centering
        \includegraphics[width=\textwidth]{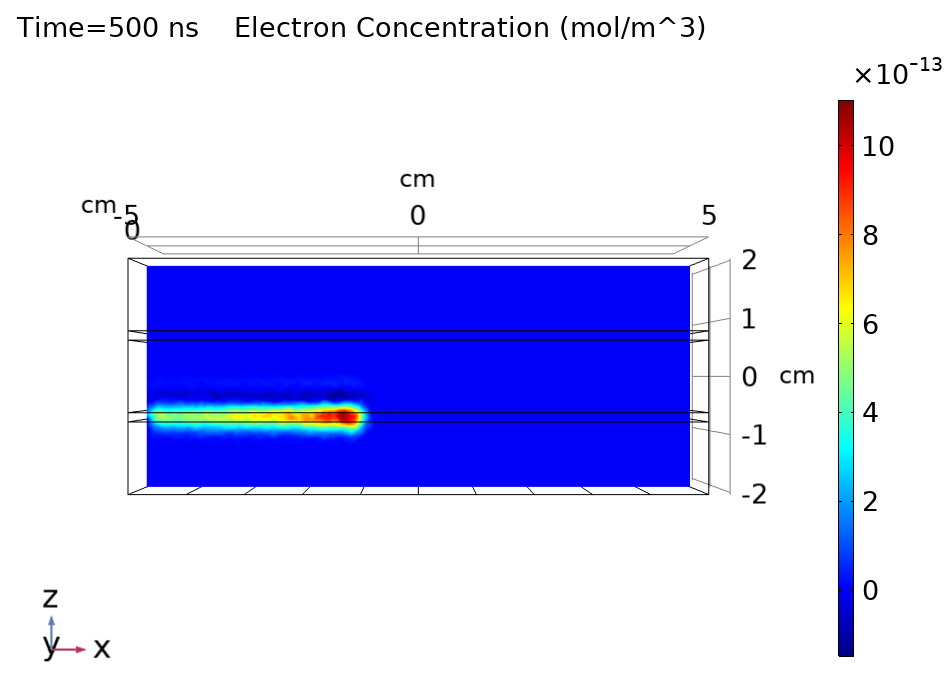}
        \caption{$t = 500$ ns}
        \label{fig:comsol_500}
    \end{subfigure}
    
    \vspace{0.5cm}
    \begin{subfigure}[b]{0.48\linewidth}
        \centering
        \includegraphics[width=\textwidth]{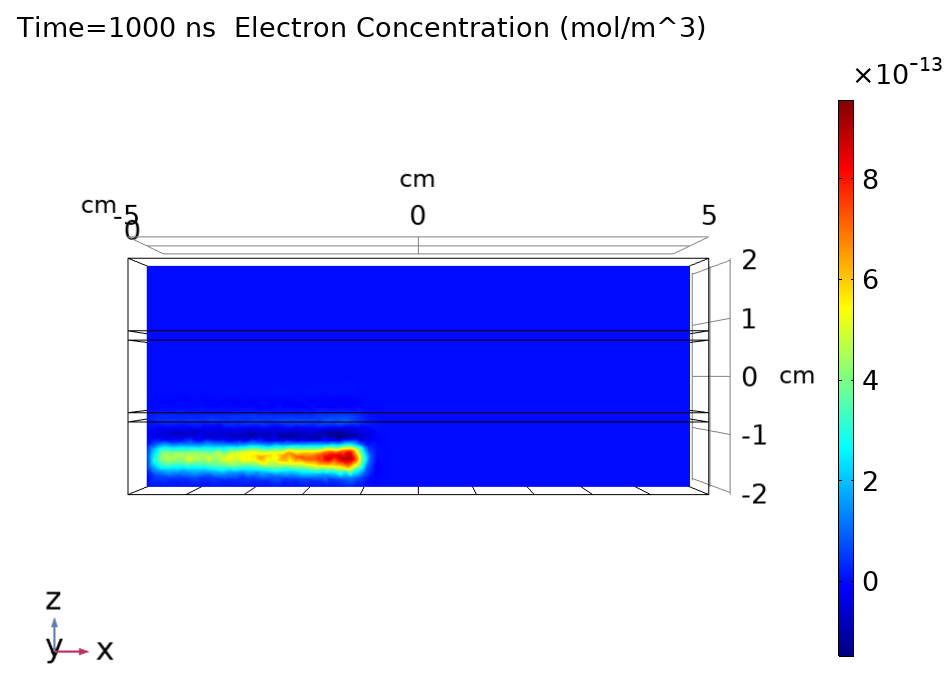}
        \caption{$t = 1000$ ns}
        \label{fig:comsol_1000}
    \end{subfigure}
    \hfill
    \begin{subfigure}[b]{0.48\linewidth}
        \centering
        \includegraphics[width=\textwidth]{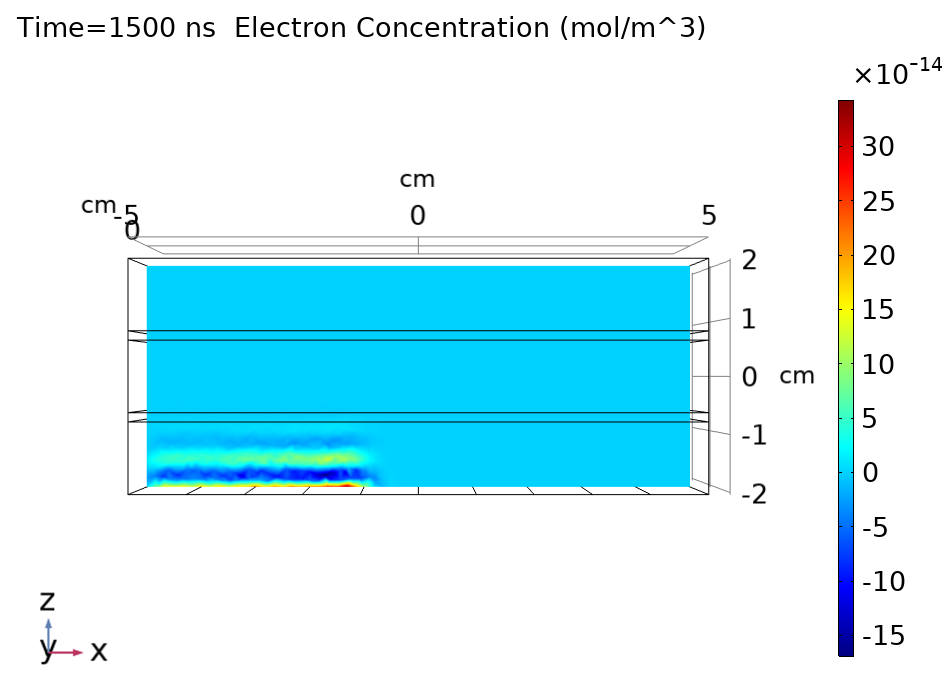}
        \caption{$t = 1500$ ns}
        \label{fig:comsol_1500}
    \end{subfigure}
    \caption{Temporal evolution of the primary electron cluster seed in Ar + CO$_2$ (90:10) at four distinct time stamps, simulated using the \textsc{COMSOL} hydrodynamic model}
    \label{sim_comsol}
\end{figure}

The drift velocity and diffusion parameters for Ar + CO$_{2}$ (90:10) and Ar + i-C$_{4}$H$_{10}$ (95:5), as obtained from \textsc{magboltz} for the given electric field, were used to simulate temporal evolution and spatial spread of primary electron cluster seed in the drift volume. One hundred $\alpha$-events were simulated for each gas mixture, generated from the source that was placed 2~cm away from the active volume at a location of $X=-7$~cm and $Y=0$~cm defining the center of the detector active volume as the origin (0,0,0). 
Figure~\ref{sim_comsol} presents the temporal evolution of the electron cloud in Ar + CO$_2$ (90:10) at four distinct time stamps as it drifted toward the readout strips. At the initial time stamp ($t = 0$~ns), the electron concentration strictly mirrored the \textsc{geant4}-produced seed cluster. As time progresses, the subsequent snapshots illustrate the steady translation of the charge cloud along the uniform drift field, accompanied by the gradual spatial broadening dictated by the transverse and longitudinal diffusion properties of the gas mixture.

\begin{figure}[htbp]
     \centering
     \begin{subfigure}[b]{0.45\linewidth}
         \centering
         \includegraphics[height=5cm,width=\textwidth]{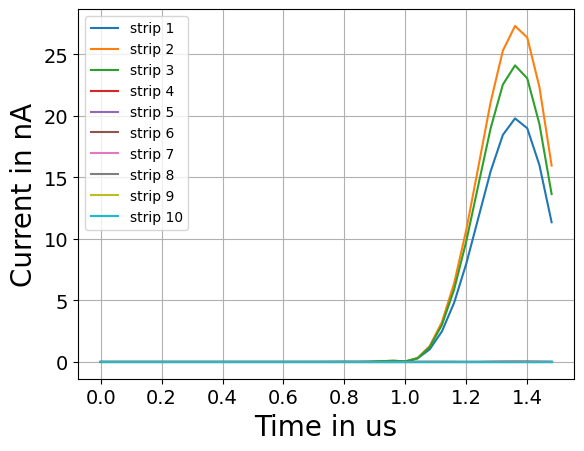}
         \caption{}
         \label{sim_current}
     \end{subfigure}
     \begin{subfigure}[b]{0.5\linewidth}
         \centering
         \includegraphics[height=5cm,width=\textwidth]{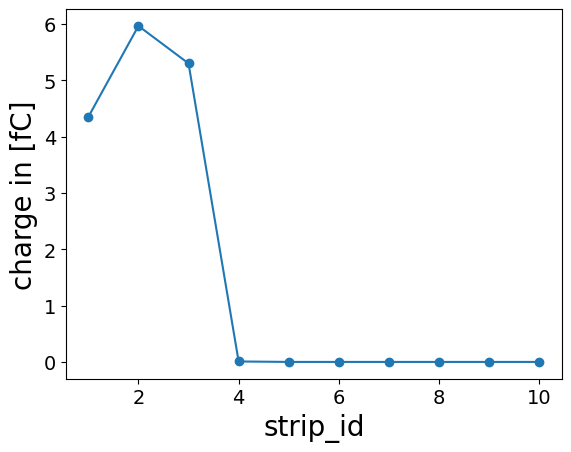}
         \caption{}
         \label{sim_charge}
     \end{subfigure}
\caption{(a) Current generated by an $\alpha$-event, and (b) primary ionization charge deposited by an $\alpha$-event, as calculated by the hydrodynamic model~\cite{DasModel2025}} 
\label{sim_track}
\end{figure}

The induced currents on the anode readout strips with dimension 10~cm $\times$ 1~cm were calculated as the electrons propagated through the drift volume, shown in figure~\ref{sim_current}. The current signals were integrated to obtain the equivalent collected charge on the readout strips, as illustrated in figure~\ref{sim_charge}.
The simulated profile exhibited a similar qualitative trend to that obtained from the experimental measurement (figure~\ref{pulse_profile}). The quantitative differences, such as the steeper charge variation observed between adjacent strips in the experimental data and the discrepancy, can be attributed to idealizations in the simulation (which excludes avalanche fluctuations and cross-talk), the coarse granularity of the readout, and localized edge field distortions near the boundary of the active region. The same kind of charge profile can be subsequently useful to determine the hit position and reconstruct the projected $\alpha$-particle track in the SAT-TPC prototype. 

\subsection{Gas Gain with $\alpha$-Particles}
\begin{figure}[htbp]
\centering
\includegraphics[height=6cm,width=0.7\textwidth]{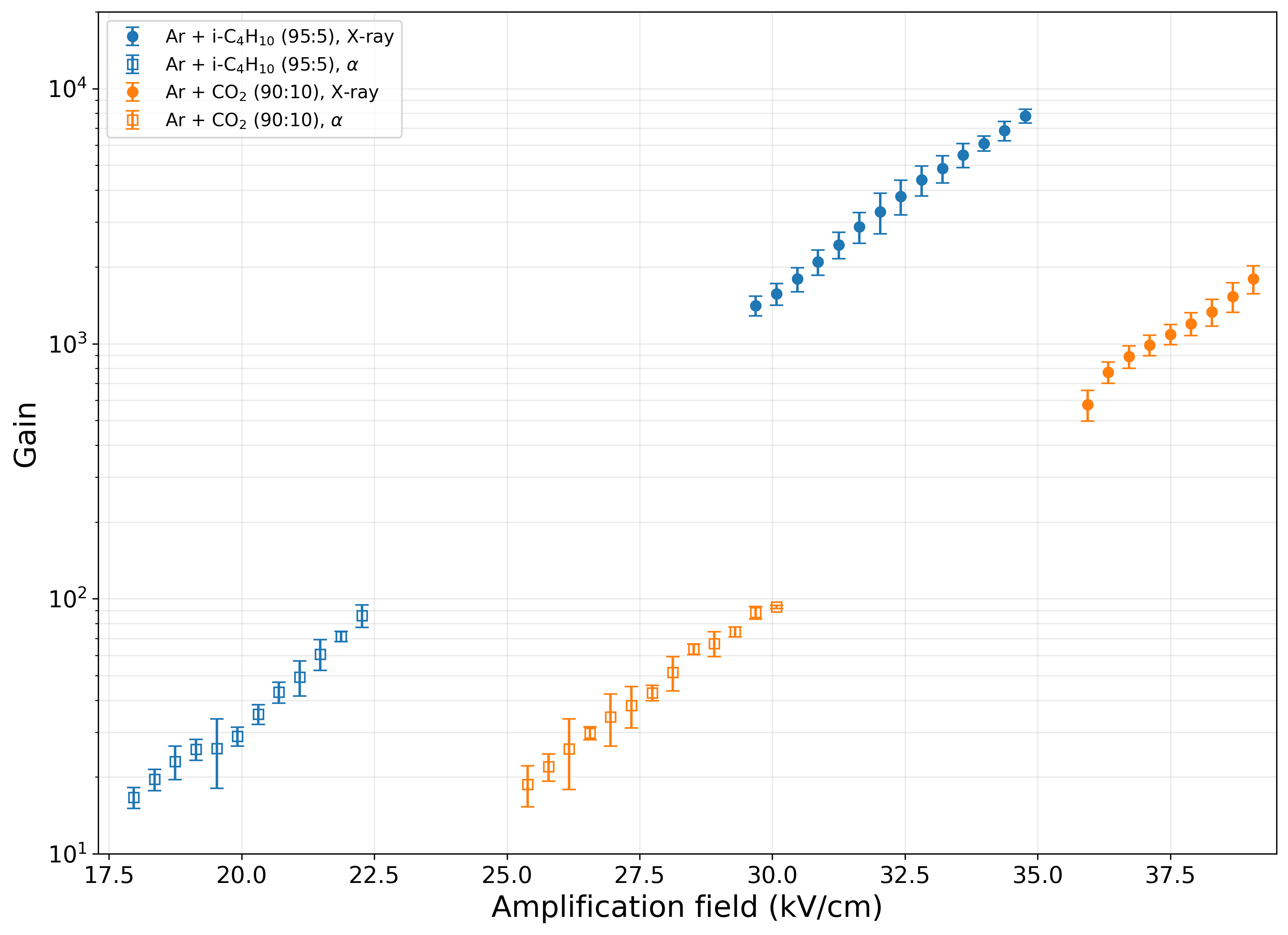}
\caption{Gas gain measured with X-rays and $\alpha$-particles in Ar + CO$_2$ (90:10) and Ar + i-C$_4$H$_{10}$ (95:5), shown at their respective experimental amplification fields.}
\label{gain_alpha}
\end{figure}

The results of the numerical simulation led to determination of the gas gain for the $\alpha$-particles. The number of primary electrons, $N_p$, was determined from the collected charge on the readout strips, simulated by the hydrodynamic model, as no multiplication was considered in the simulation (illustrated in figure~\ref{sim_charge}). The collected charge, $Q_{col}$, in terms of ADC channel was obtained from the integrated ADC counts (shown in figure~\ref{pulse_profile}) which can be converted to charge using the calibration data. 
The gas gain of the Micromegas setup, estimated for 5.48~MeV $\alpha$-particles in Ar + CO$_2$ (90:10) and Ar + i-C$_4$H$_{10}$ (95:5), is depicted in figure~\ref{gain_alpha}. The prototype provided effective gains ranging from 10 to 100 over amplification-field ranges of 18--22.5~kV/cm for
Ar + i-C$_4$H$_{10}$ (95:5) and 25.5--30~kV/cm for
Ar + CO$_2$ (90:10). For both gas mixtures, the measured
gain increased with increasing amplification field. Since the measurements were performed over different amplification-field ranges and the experimental uncertainties are non-negligible, no quantitative comparison of the absolute gain between the two mixtures is made from these data. The absolute gain values follow trends similar to those reported for Micromegas operated in Ar-based mixtures~\cite{Dafni2009}, but remain below the higher gains obtained in configurations employing additional pre-amplification stages or operated at larger amplification fields~\cite{Cortesi2016,Koshchiy20}. The Micromegas was operated at lower amplification fields for the
$\alpha$-particle measurements than for the X-ray measurements because
of the substantially higher primary ionization charge density produced
by the $\alpha$-particles, thereby reducing the probability of
micro-discharges. To compare the gain behavior in the two operating regimes, the measured X-ray and $\alpha$-particle gains are shown together in figure~\ref{gain_alpha} at their respective experimental amplification fields. The comparison highlights the distinct operating-field ranges of the two measurements while preserving the observed field-dependent increase in gain for both gas mixtures.

\subsection{Energy Resolution with $\alpha$-Particles}
\begin{figure}[htbp]
     \centering
     \begin{subfigure}[b]{0.45\linewidth}
         \centering
         \includegraphics[height=5cm,width=\textwidth]{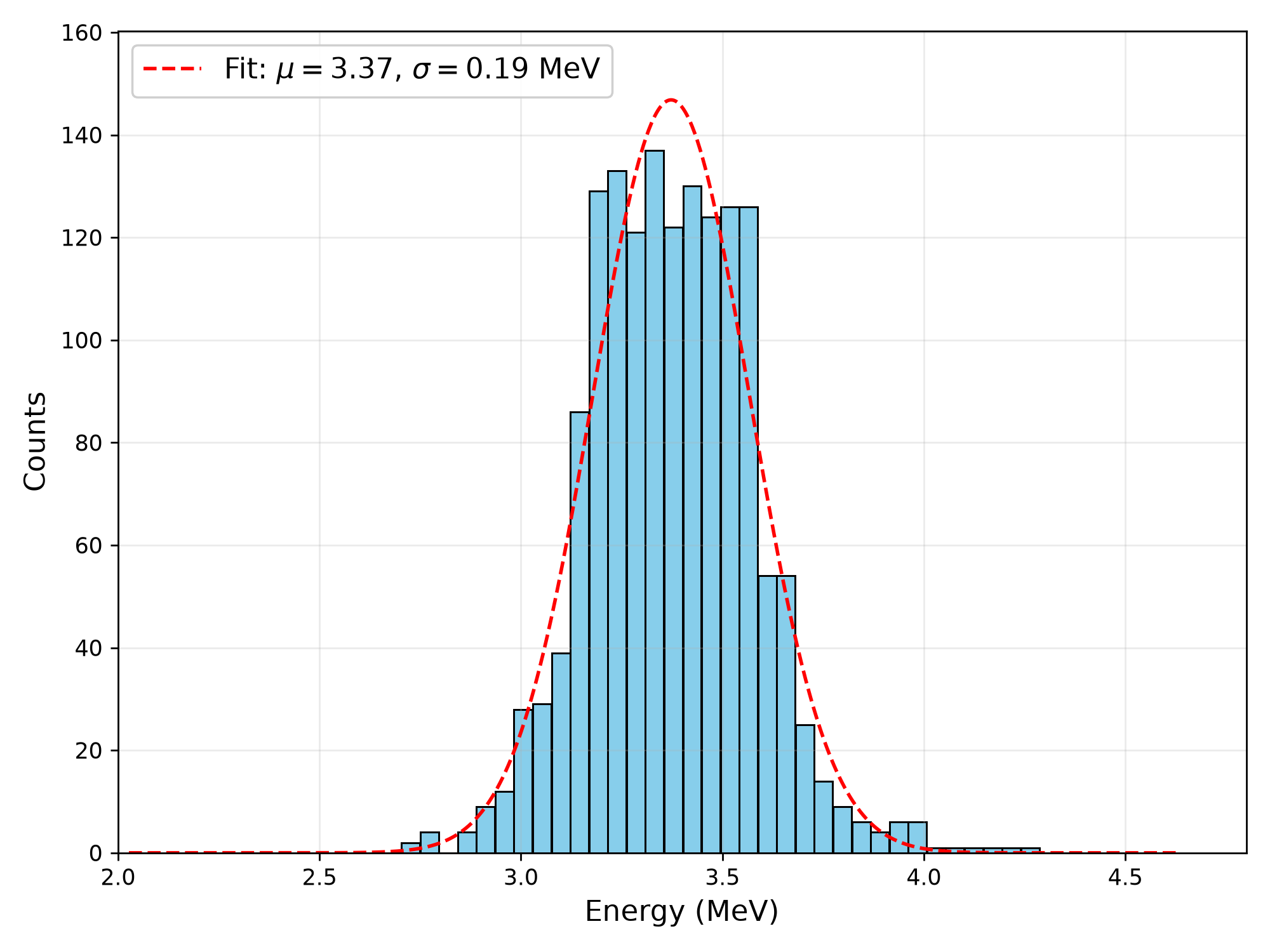}
         \caption{}
         \label{energy_ic4h10}
     \end{subfigure}
     \begin{subfigure}[b]{0.5\linewidth}
         \centering
         \includegraphics[height=5cm,width=\textwidth]{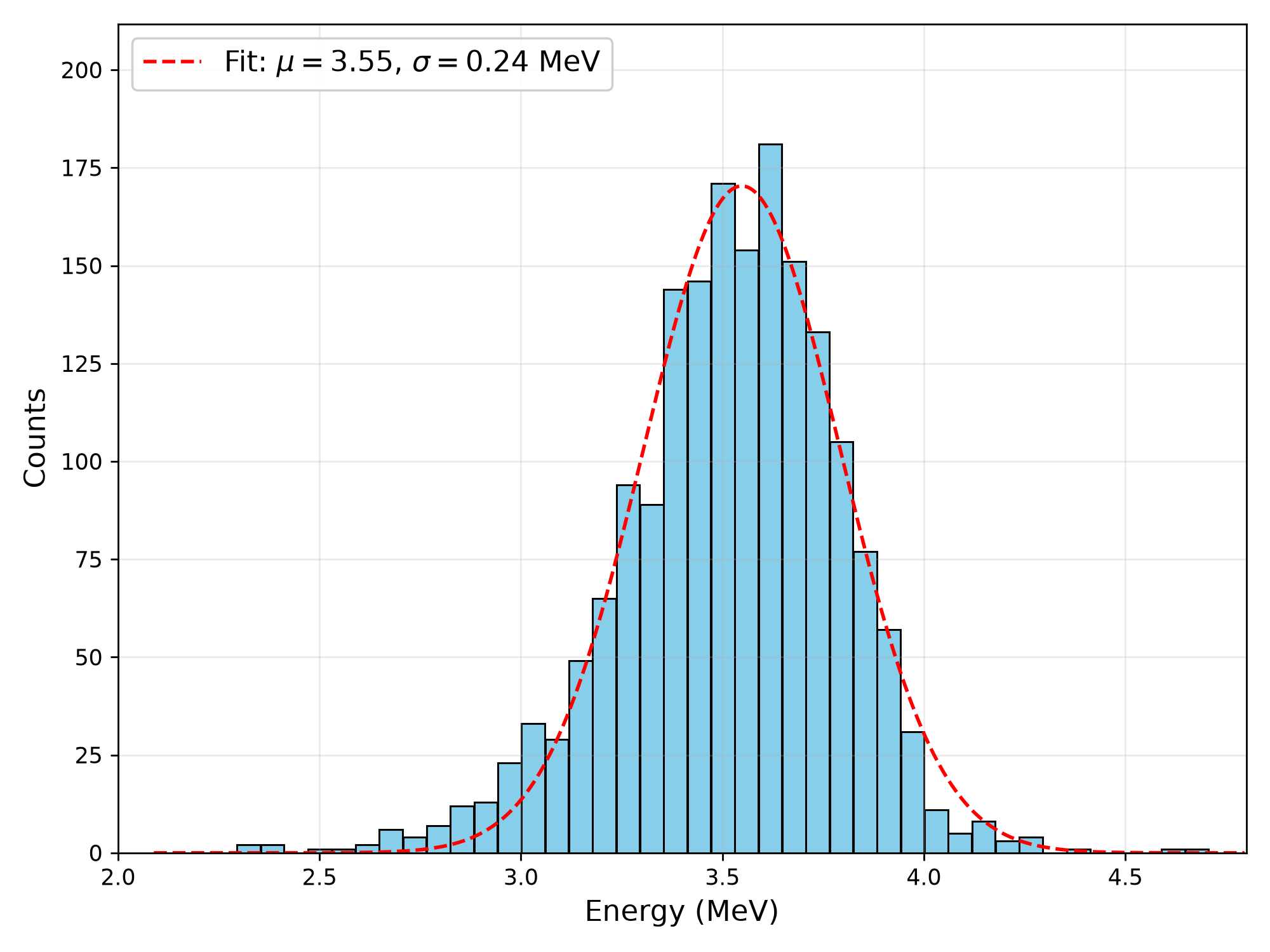}
         \caption{}
         \label{energy_co2}
     \end{subfigure}
   \caption{Spectrum of the energy deposited by $\alpha$-particles in (a) Ar + i-C$_4$H$_{10}$ (95:5) and (b) Ar + CO$_2$ (90:10)}  
        \label{energy}
\end{figure}

The energy resolution for $\alpha$-particles was studied for both the gas mixtures. While taking data, the micro-mesh voltage was set at -340~V (yielding gain $\sim29$) for Ar + CO$_2$ (90:10) and -240~V (yielding gain $\sim23$) for Ar + i-C$_4$H$_{10}$ (95:5). The corresponding drift cathode voltages were -1360~V and -960~V, respectively. 
The energy deposited by the $\alpha$-particles was reconstructed by summing the calibrated charge obtained from the charge-deposit profile of the readout strips. To obtain the final energy values, the total collected charge for each event was scaled by the effective gas gain determined for the respective gas mixture and multiplied by the mean energy per ion pair (W-value) to account for the primary ionization energy. The effective $W$-values were consistently determined from the
HEED-simulated mean primary-electron yields discussed in
Section~\ref{gg_fe}. For the 5.9~keV X-rays, the simulated mean values of
$N_p=211$ and $218$ correspond to effective $W$-values of
approximately 28.0~eV for Ar + CO$_2$ (90:10) and 27.1~eV for
Ar + i-C$_4$H$_{10}$ (95:5), respectively. These mixture-specific
values were therefore used for the reconstruction of the
$\alpha$-particle energy. The reconstructed peak positions are approximately 3.37~MeV for
Ar + i-C$_4$H$_{10}$ (95:5) and 3.55~MeV for Ar + CO$_2$ (90:10).
These values are close to the simulated deposited energy of
3.43~MeV, within the experimental energy resolution and the
systematic uncertainties associated with the gain and charge
calibration.
The energy spectra obtained from the sum of the strip charges for both the gas mixtures are shown in figure~\ref{energy}.
The measured energy resolutions are $\sigma$/$\mu$ = 5.8\% for Ar + i-C$_4$H$_{10}$ (95:5) and $\sigma$/$\mu$ = 6.9\% for Ar + CO$_2$ (90:10), corresponding to FWHM values of approximately 13.7\% and 16.2\%, respectively. The improved resolution observed for Ar + i-C$_4$H$_{10}$ (95:5)
is consistent with reduced electron diffusion and enhanced Penning
transfer in this gas mixture. The relatively coarse 1 cm strip pitch currently limits the precision of the energy reconstruction. While simple charge summation across multiple strips inherently introduces additional electronic noise, implementing a finer readout granularity would provide sufficient spatial sampling of the track to extract the energy through range analysis and fitting of the Bragg peak profile. Such track profiling methods have been shown to significantly improve final energy resolution in active targets compared to total charge summation~\cite{Suzuki2013}. The present results fall within the range typically observed for bulk Micromegas readouts in AT-TPC and TexAT-type systems.

%and are naturally broader than the $\sim$2\% FWHM achieved in high-pressure microbulk detectors~\cite{Dafni2009}. Overall, the measured resolutions are consistent with expectations for a 128~$\mu$m bulk-Micromegas with 1~cm readout strip pitch, operated at atmospheric pressure. 

\section{Summary \& Conclusion}
\label{Con}
A bulk Micromegas was characterized in an experimental setup with a 4 cm drift volume by evaluating its response to X-rays and $\alpha$-particles from radioactive sources, $^{55}$Fe and $^{241}$Am, respectively.
The Micromegas was operated with Ar + CO$_2$ (90:10) and Ar + i-C$_4$H$_{10}$ (95:5), providing stable operation in the maximum-transparency region, and achieved gains of about 2000 and $10^4$, respectively, for X-rays. For $\alpha$-particles, the Micromegas delivered an effective gas gain in the range of 10 to 100 in those gas mixtures. The energy resolution for the X-ray was found to be $\sigma/\mu$ = 9.7\% in Ar + CO$_2$ and $8.5\%$ in Ar + i-C$_4$H$_{10}$, while those were $6.9\%$ and $5.8\%$ for the $\alpha$-particles, consistent with expectations for a 128~$\mu$m bulk
Micromegas. 
The $\alpha$-track profile obtained from the readout strip signals showed a qualitative consistency with the primary ionization charge profile produced 
with the numerical simulation based on hydrodynamic approach, carried out using \textsc{geant4} and \textsc{comsol}, validating the scope of the numerical model.
%and demontrated that the SAT-TPC prototype equipped with the bulk Micromegas could recover the projected track
%length and direction satisfactorily within its limited design capabilities.
The observed energy resolution demonstrates that the bulk-Micromegas provides a robust amplification stage for the SAT-TPC prototype. Furthermore, the spatial charge collection profile indicates that, upon upgrading to a higher granularity readout, the system will possess the spatial resolution necessary for three dimensional particle tracking.
The future work will focus on improving the energy resolution and tracking granularity by employing finer readout segmentation, exploring operation at lower pressure, and performing in-beam tests to validate the Micromegas-based operation of SAT-TPC prototype performance.
\section*{Acknowledgments}
The author, P. K. Das, would like to thank Mr. Nilanjan Biswas of SINP for his technical help in assembling the setup and his seniors Drs. A. Jash, T. Dey, V. Kumar and S. Das for their valuable suggestions and advice for the experimental work and data analysis. The authors P. K. Das and N. Majumdar acknowledge the financial and infrastructural support provided by their institute.

\end{document}